\documentclass[reprint,amsmath,amssymb,aps,floatfix,twocolumn,longbibliography,prd]{revtex4-2}
\usepackage{subfigure}
\usepackage[colorlinks=true]{hyperref}
\usepackage{graphicx}%
\usepackage{dcolumn}%
\usepackage{bm}\allowdisplaybreaks
\usepackage{orcidlink}

\usepackage{amsmath}
\usepackage{amssymb}
\usepackage{mathrsfs}

\usepackage{bibentry}

\begin{document}

\title{Gross-Neveu-Yukawa theory of $\text{SO}(2N)\rightarrow \text{SO}(N) \times \text{SO}(N)$ spontaneous symmetry breaking}

\author{SangEun Han\,\orcidlink{0000-0003-3141-1964}
}
\affiliation{Department of Physics, Simon Fraser University, Burnaby, British Columbia, Canada V5A 1S6}

\author{Igor F.~Herbut\,\orcidlink{0000-0001-5496-8330}
}
\affiliation{Department of Physics, Simon Fraser University, Burnaby, British Columbia, Canada V5A 1S6}

\date{\today}

\begin{abstract}
We construct and study the relativistic Gross-Neveu-Yukawa field theory for the $\text{SO}(2N)$ real symmetric second-rank tensor order parameter coupled to $N_f$ flavors of $4N$-component Majorana fermions in 2+1 dimensions. Such a tensor order parameter unifies all Lorentz-invariant mass-gap orders for $N$ two-component Dirac fermions in two dimensions except for the $\text{SO}(2N)$-singlet anomalous quantum Hall state. The value $N_f=1$ corresponds to the canonical Gross-Neveu model. Within the leading-order $\epsilon$-expansion around the upper critical dimension of $3+1$ the field theory exhibits a critical fixed point in its renormalization group flow which describes spontaneous symmetry breaking to $\text{SO}(N)\times \text{SO}(N)$ for the number of flavors of Majorana fermions higher than a critical value $N_{f,c2}\approx 2N$. For $N_{f, c1}< N_f < N_{f,c2}$ , with $N_{f,c1} \approx N$ the critical fixed point resides in the unstable region of the theory where the effective potential is unbounded from below, whereas for $N_f < N_{f,c1}$ there is no real critical fixed point, and the flow runs away. In either case, for $N_f < N_{f,c2}$ the transition should become fluctuation-induced first-order, and we discuss the dependence of its size on the parameters $N$ and $N_{f}$ in the theory. One-loop critical exponents for the universality class at $N_{f, c2}< N_f $ are computed and the flow diagram in various regimes is discussed.
\end{abstract}

\maketitle

\section{Introduction}

The Ising symmetry-breaking transition in the Gross-Neveu (GN) model \cite{gross, zinnjustinbook} in 2+1 dimensions has been studied by the large-$N$ expansion and the conformal bootstrap \cite{rosenstein, vasilev, gracey1, gracey2, erramilli}, and is believed to be reasonably well understood. At this continuous transition the \text{SO}($2N$) flavor symmetry of the GN model remains preserved, and the discrete Ising symmetry becomes spontaneously broken. Important role in our present understanding of the GN universality class is played by the closely related Gross-Neveu-Yukawa (GNY) description \cite{zinnjustin}, in which the real Ising order parameter is explicitly included,  with its own relativistic dynamics, Yukawa coupling to fermions, and a single contact self-interaction. The GNY field theory then has two interaction couplings that both would  become marginal in 3+1 dimensions, which facilitates the standard $\epsilon$-expansion around the upper spatial critical dimension. Extensions of the original GNY theory have been proposed  \cite{herbutprl, hjr} and studied as  possible descriptions of the Dirac semimetal - Mott insulator transition in two-dimensional electronic systems such as graphene and $d$-wave superconductors \cite{hjv, mihaila, zerf, ihrig, vojta, huh, schwab, scherer, torres, herbutfisherbook}.

The GN model for $N$ flavors of two-component Dirac fermions in 2+1 dimensions, however, also possesses the $\text{SO}(2N)$ symmetry which can become spontaneously broken down to $\text{SO}(N) \times \text{SO}(N)$ \cite{so8}.  At this separate transition one (or more) of the components of the order parameter that transform as second-rank real irreducible tensor under $\text{SO}(2N)$ becomes nonzero. The value of $N=4$ is relevant to graphene, for example, and the second-rank tensor unifies all order parameters that play the role of relativistic mass for Dirac fermions, except one, which is a singlet under the  $\text{SO}(2N)$. It was argued recently \cite{han} that fluctuations may turn this transition into first order, and that the continuous transition obtains only when the $\text{SO}(2N)$ symmetry of the GN model is further enlarged into $\text{SO}(2N) \times \text{SO}(N_f)$, and $N_f$, which may be understood as the number of flavors of $2N$-component Dirac fermions, is taken sufficiently large. At $N_f=1$, which corresponds to the original GN model, the critical fixed point that exists at large $N_f$ was found to lose its diverging susceptibility \cite{han}, and this way rendered unphysical. This was interpreted as an indication of the fluctuation-induced first-order transition.

In this paper we scrutinize further the spontaneous breaking of the $\text{SO}(2N)$ symmetry of the GN model by considering a suitably formulated GNY field theory, and by using the expansion around its upper critical dimension of 3+1. Since the order parameter is a $2N$-dimensional, real, traceless symmetric matrix, and $N\geq 2$, the bosonic part of the action contains {\it two} distinct quartic self-interaction terms. The order parameter field theory without any coupling to the fermions is known to exhibit only the runaway flow in the renormalization group (RG) and have no stable fixed points \cite{delfino, pelissetto}. We find, however, that a sufficiently large number of relativistic fermions Yukawa-coupled to the bosonic matrix field  produces a stable fixed point in the RG flow and leads to a continuous transition which belongs to a new universality class. This is in agreement with our previous study done directly on the GN model, when extended to values $N_f >1$ \cite{han}. We determine the critical exponents and the mass-gap ratio at this fixed point, discuss the conditions for its existence, and connect the RG ``time" for the runaway flows to the size of the first order transition.

The paper is organized as follows. In the next section we begin by reviewing the GN model and the standard (Ising) GNY field theory. In Sec.~\ref{sec:symmetric_rep} we discuss the emergence of the $\text{SO}(2N)$ symmetry, define (``real") Majorana fermions, and recognize the order parameter as the real irreducible symmetric tensor.  In Sec.~\ref{sec:tensor_field_theory} we recall and discuss salient features of the field theory for the general real symmetric $\text{SO}(M)$ tensor order parameter alone, decoupled from fermions. The GNY theory of the matrix order paramater Yukawa-coupled to the Majorana fermions is finally presented and analyzed in Sec.~\ref{sec:GNY} and \ref{sec:GNY_Nf}. Summary and further comments are given in Sec.~\ref{sec:discussion}.

\section{Ising GNY field theory}

We begin by briefly reviewing the phase transition in the standard GN model in 2+1 dimensions. We define its action as
\begin{align}
\mathcal{S}_{\text{GN}}%
={}&\int d\tau d^{d}x\;[\psi^{\dagger}(\mathbb{I}_{N}\otimes(\mathbb{I}_{2}\partial_{\tau}-i\sigma_{1}\partial_{1}-i\sigma_{3}\partial_{2}))\psi\notag\\&\quad\quad\quad\quad\quad+\bar{g}(\psi^{\dagger}(\mathbb{I}_{N}\otimes\sigma_{2})\psi)^{2}],
\label{eq:GN_action}
\end{align}
where $\sigma_{i}$ are the Pauli matrices, $\mathbb{I}_{N}$ is the $N\times N$ identity matrix, and $\psi=\psi(x,\tau)$ is a $2N$-component Grassmann (complex Dirac) field, $\tau$ is the imaginary time and $d=2$. By performing the standard Wilson RG calculation by integrating out fermions with momenta between $\Lambda/b$ and $\Lambda$ and with all frequencies, to the leading order in the interaction $\bar{g}$  one finds the following (one-loop) beta-function,
\begin{align}
\frac{dg}{d\ell}=- g-(N - 1)g^{2}+\mathcal{O}(g^3),\label{eq:GN_RG}
\end{align}
where $g=\bar{g}\Lambda/(2\pi)$, $\Lambda$ is the ultraviolet cutoff on the momentum integration, and $\ell=\ln b$. The beta-function has a critical fixed point at the negative $g_{c1}^{*}=1/(1-N)$. If $g< g_{c1}^{*}$, $g\rightarrow-\infty$ under RG, and  $\langle\psi^{\dagger}(\mathbb{I}_{N}\otimes\sigma_{2})\psi\rangle\neq 0$, and the fermion becomes gapped. If $0>g>g_{c1}^{*}$, $g\rightarrow0$ under RG,  i.\,e.~it flows to the  non-interacting Gaussian fixed point. When $g$ is small and positive, the RG flow is also towards the non-interacting Gaussian fixed point. The coefficient of the two-loop term in the beta-function is also of order $N$ at large $N$, and provides the next-order term in the systematic $1/N$-expansion, which at the time of writing has been pushed to the order $1/N^3$ \cite{gracey1}. Note that the order parameter is a singlet under the flavor $\text{U}(N)$ symmetry of the GN model, under which $\psi \rightarrow (U\otimes \mathbb{I}_{2}) \psi$.

An alternative point of view at the transition in the GN model may be taken by first Hubbard-Stratonovich transforming the quartic term and then supplying the relativistic dynamics to the (real) Hubbard-Stratonovich field. The dynamics of the order parameter (Hubbard-Stratonovich) field would emerge from the integration of high-energy Dirac femions. This way one finds the GNY theory for the Dirac fermions and mass order-parameter Ising (real) field in the form:
\begin{widetext}
\begin{align}
    \mathcal{S}_{\text{GNY}}={}&\int d\tau d^{d}x\;\left[\psi^{\dagger}(\mathbb{I}_{N}\otimes( \mathbb{I}_{2}\partial_{\tau}-iv_{F}\sigma_{1}\partial_{1}-iv_{F}\sigma_{3}\partial_{2}))\psi+\bar{g}\varphi(\psi^{\dagger}(\mathbb{I}_{N}\otimes\sigma_{2})\psi)+\frac{1}{2}[(\partial_{\tau}\varphi)^{2}+v_{B}^{2}(\nabla\varphi)^{2}]+\frac{1}{4}\bar{\lambda}\varphi^{4}\right],
\end{align}
\end{widetext}
where $v_{F/B}$ are the velocities of the fermion and boson fields, respectively, which we for generality allow to be different. $\bar{\lambda}$ is the self-interaction coupling constant for the order parameter, and we tuned the coefficient of the quadratic term $\varphi^2 $, $r$,   to zero. The one-loop RG flow equations are now given by \cite{zinnjustin,Han2018}
\begin{align}
\frac{dy}{d\ell}={}&\alpha_{g}\left(\frac{y^{2}}{3(1+y)^{2}}+\frac{N(1+y)}{8}\right)y(1-y),\\
\frac{d\alpha_{g}}{d\ell}={}&\epsilon\alpha_{g}-\alpha_{g}^{2}\left(\frac{N}{4}+\frac{y^{2}(1+2y)}{(1+y)^{2}}+\frac{y^{2}(1-y)}{(1+y)^{2}}\right),\\
\frac{d\lambda}{d\ell}={}&\epsilon\lambda-\alpha_{g}\lambda\frac{N}{2}+\alpha_{g}\lambda\frac{3N(1-y^{2})}{8}-\frac{9}{4}\lambda^{2}+\alpha_{g}^{2}\frac{Ny^{3}}{2},
\end{align}
where $\epsilon=3-d$,
\begin{align}
y=\frac{v_{F}}{v_{B}},\; \alpha_{g}=\frac{\bar{g}^{2}S_{d}}{(2\pi)^{d}v_{F}^{d}\Lambda^{d-3}},\; \lambda=\frac{\bar{\lambda}S_{d}}{(2\pi)^{d}v_{B}^{d}\Lambda^{d-3}},
\end{align}
and $S_{d}=2\pi^{d/2}/\Gamma(d/2)$ is the area of the sphere in $d$ dimensions. 
$dy/d\ell$ has a single stable fixed point value at $y=1$, so, as usual \cite{roy}, there is an emergent Lorentz symmetry in the infrared. At $y=1$, the remaining beta-functions simplify into
\begin{align}
\frac{d\alpha_{g}}{d\ell}={}&\epsilon \alpha_{g}-\alpha_{g}^{2}\frac{(N+3)}{4},\\
\frac{d\lambda}{d\ell}={}&\epsilon \lambda-\frac{9\lambda^2 }{4}-\alpha_{g}\lambda\frac{N}{2}+\alpha_{g}^{2}\frac{N}{2}.
\end{align}
The one-loop beta-function for the Yukawa coupling decouples and yields the attractive fixed point at $\alpha_{g}^{*}=4\epsilon/(N+3)$. At $\alpha_{g}=\alpha_{g}^{*}$, the remaining beta-function for the self-interaction $\lambda$ becomes
\begin{align}
\frac{d\lambda}{d\ell}={}&\frac{3-N}{3+N}\epsilon \lambda-\frac{9\lambda^{2}}{4}+\frac{8N\epsilon^{2}} {(3+N)^{2}}.
\end{align}
The attractive real fixed point value of $\lambda$ then exists at all values of $N$, and equals
\begin{align}
\lambda^{*}=\frac{2\epsilon}{9}\frac{(3-N+\sqrt{9+66N+N^{2}})}{(3+N)}.
\end{align}
$\lambda^{*}$ is also positive, so the order parameter effective potential is bounded from below, and the theory is stable. Together these results imply a second-order phase transition \cite{herbutbook}.

Since $m_{\varphi}^{2}=-2r$ and $m_{\psi}^{2}=-\frac{g^{2}r}{\lambda}$ are the squares of the masses of the order parameter and Dirac fermions in the ordered phase, respectively, the dimensionless mass-ratio can be written as
\begin{align}
\mathcal{R}_{G}=\frac{m_{\varphi} ^2 }{m_{\psi} ^2 }={}&\frac{2\lambda^{*}}{\alpha_{g}^{*}}=\frac{1}{9}(3-N+\sqrt{9+66N+N^{2}}).\label{eq:massgap_GN}
\end{align}
We note that  $ \mathcal{R}_{G} \rightarrow 4$ as $N\rightarrow \infty$, reflecting the simple composite nature of the boson in this limit.

The correlation length exponent at the fixed point is similarly given by
\begin{align}
\nonumber
\nu^{-1}={}& \left( 2-\frac{3\lambda^* }{4}-\frac{N\alpha_{g} ^* }{4}  \right) \\
=& 2-\frac{\epsilon}{6} \left( \frac{5N+3}{N+3}+\frac{\sqrt{9+66N+N^{2}})}{(3+N)} \right).
\end{align}

One can also compute the anomalous dimensions of the fermion and boson fields, $\eta_{\psi}$ and $\eta_{\varphi}$:
\begin{align}
\eta_{\psi}={}&\frac{\epsilon}{2(N+3)}
,\quad
\eta_{\varphi}=\frac{N\epsilon}{N+3}.
\end{align}
By substituting $N\rightarrow 2N$ one recovers the correct results for the universal quantities close to $d=3$, where one has $N$ copies of four-component Dirac spinors, and the requisite four-dimensional Dirac matrices instead of Pauli matrices \cite{zinnjustin,Han2018}.

Finally, let us observe in passing that the GN and GNY descriptions of the phase transition match neatly in the limit $N\gg 1$. In this limit the fixed point is at $\alpha_{g}^{*}=4\epsilon/N$ and $\lambda^{*} = 8\epsilon/N$,  to the leading order in $1/N$. This yields $\nu= 1 + \mathcal{O}(1/N)$, $\eta_\varphi = 1$, the correct leading order results for $\epsilon=1$.

\section{SO($2N$) symmetry and the tensor representation}\label{sec:symmetric_rep}

In the previous work \cite{so8} by a combination of several Fierz identities it was shown that the standard GN interaction term can be rewritten in a suggestive way as
\begin{align}
-(N&+1)(\psi^{\dagger}(\mathbb{I}_{N}\otimes\sigma_{2})\psi)^ {2}  \notag\\
 ={}& (\psi^{\dagger}(G_{a}\otimes\sigma_{2})\psi)^{2}+(\psi^{\dagger}(S_{b}\otimes\sigma_{2})\psi^{*})(\psi^{\intercal}(S_{b}\otimes\sigma_{2})\psi), \label{eq:fierz_GN}
\end{align}
where $G_{a}$ are the generators of the SU($N$) in the fundamental representation ($a=1,\cdots,N^{2}-1$), $\text{Tr} [G_{a1} G_{a2}]=N \delta_{a1a2}$, and $S_{b}$ are  linearly independent, real, symmetric $N$-dimensional matrices $b=1,\cdots,N(N+1)/2$, and $\text{Tr} [S_{b1} S_{b2}]= N \delta_{b1b2}$.

Another similar identity can also be derived \cite{so8,han},
\begin{align}
&-3N(\psi^{\dagger}(\mathbb{I}_{N}\otimes\sigma_{2})\psi)^{2}
=(\psi^{\dagger}(\mathbb{I}\otimes\sigma_{2}\sigma_{i})\psi)^{2} +  \notag\\
&\quad(\psi^{\dagger}G_{a}\otimes\sigma_{2}\sigma_{i})\psi)^{2}+
(\psi^{\dagger}(A_{c}\otimes\sigma_{2}\sigma_{i})\psi^{*})(\psi^{\intercal}(A_{c}\otimes\sigma_{2}\sigma_{i})\psi),
\end{align}
where $A_{c}$ are linearly independent, imaginary, antisymmetric, traceless $N$-dimensional matrices ($c=1,\cdots,N(N-1)/2$), and $\text{Tr}[A_{c1}A_{c2}]=N\delta_{c1c2}$. Using these two identities the GN model in  Eq.~\eqref{eq:GN_action} can be rewritten as
\begin{widetext}
\begin{align}
\bar{\mathcal{S}}_{\text{GN}}={}&\int d\tau d^{2}x\;\left[\psi^{\dagger}(\mathbb{I}_{N}\otimes(\mathbb{I}_{2}\partial_{\tau}-i\sigma_{1}\partial_{1}-i\sigma_{3}\partial_{2}))\psi\notag\right.\\
&+\bar{g}_{1}(\psi^{\dagger}(\mathbb{I}_{N}\otimes\sigma_{2})\psi)^{2}-\frac{\bar{g}_{2}}{(N+1)}[(\psi^{\dagger}(G_{a}\otimes\sigma_{2})\psi)^{2}+(\psi^{\dagger}(S_{b}\otimes\sigma_{2})\psi^{*})(\psi^{\intercal}(S_{b}\otimes\sigma_{2})\psi)] \notag\\
&\left.-\frac{\bar{g}_{3}}{3N}[(\psi^{\dagger}(\mathbb{I}_{N}\otimes\sigma_{2}\sigma_{i})\psi)^{2}+(\psi^{\dagger}(G_{a}\otimes\sigma_{2}\sigma_{i})\psi)^{2}+(\psi^{\dagger}(A_{c}\otimes\sigma_{2}\sigma_{i})\psi^{*})(\psi^{\intercal}(A_{c}\otimes\sigma_{2}\sigma_{i})\psi)]\right]
\label{eq:g1g2g3}
\end{align}
\end{widetext}

The three four-fermion interactions in Eq.~\eqref{eq:g1g2g3} are actually identical, so that defining $\bar{g}=\bar{g}_{1}+\bar{g}_{2}+\bar{g}_{3}$ leads to the same beta-function as in  Eq.~\eqref{eq:GN_RG}. If we introduce $N_f$ copies of the $2N$-component Dirac fermion the three interaction terms become linearly independent for $N_f >1$. One-loop RG in the large-$N_f$ limit then leads to three distinct critical fixed points, which correspond to mean-field transitions into order parameters that transform as singlet, symmetric tensor, and antisymmetric tensor under the symmetry group $\text{SO}(2N)$, which we will discuss shortly. Interested reader is referred for details of this analysis to our previous paper for details of this analysis\cite{han}. Here we focus exclusively on the transition into the symmetric tensor order parameter, and ignore other possibilities.

One of the fixed points that the one-loop RG at large-$N_f$ and large-$N$ finds is the  ``symmetric-tensor" critical point  at $g_{2}^{*} \sim 1/N_f$, and $N_f ^4 g_{1}^{*}\sim N_f ^2 g_{3}^{*} \sim 1 $ \cite{han}. The analysis of susceptibilities shows that for $g_{2}>g_{2}^{*}$ Dirac fermions develop a mass gap in which $ \langle \psi^{\dagger}(G_{a}\otimes\sigma_{2})\psi \rangle$, or $\langle \psi^{\dagger}(S_b\otimes\sigma_{2})\psi^* \rangle$, or some of their (restricted) linear combinations, become finite. By reducing the parameter $N_f$ towards its original value of $N_f=1$ in the GN model, however, below certain critical value no susceptibility is found to diverge near the ``symmetric tensor" critical point, and the critical point is this way rendered unphysical. Furthermore, although the critical point survives the limit $N_f \rightarrow 1$,  at $N_f=1$ it becomes equivalent to the Gaussian fixed point. This ``mutation" of the critical point from physical at $N_f\gg 1$ to redundant at $N_f=1$ was interpreted as the sign of the $\text{SO}(2N)$-symmetry-breaking in the GN model becoming first order.

To try to shed further light on this phenomenon and attempt to understand it in more familiar terms such as fixed-point collision \cite{kubota, kaveh, gies, kaplan, qed31, gukov, gorbenko} here we turn to the GNY formulation.

Let us therefore take $\bar{g}_1 = \bar{g}_3 =0$ in Eq.~\eqref{eq:g1g2g3} and Hubbard-Stratonovich transform the $\sim \bar{g}_2$ interaction term, assuming $\bar{g}_2 >0$. In terms of the  Majorana fermions one then finds the following Lagrangian \cite{so8}:
\begin{align}
\mathcal{L}={}&\frac{1}{2}\phi^{\intercal}(\mathbb{I}_{2N}\otimes(\sigma_{0}\partial_{\tau}-i\sigma_{1}\partial_{1}-i\sigma_{3}\partial_{2}))\phi\notag\\
&+ \frac{1}{2}\phi^{\intercal}(S\otimes\sigma_{2})\phi + \frac{N+1}{8 N\bar{g}_2} \text{Tr}[S^{2}] .\label{eq:HSM_lagrangian}
\end{align}

Majorana (``real") $4N$-component fermion is defined as $\phi^{\intercal}=(\psi_{1} ^{\intercal} ,\psi_{2} ^{\intercal} )$,  where $\psi_{1,2}$ are related to Dirac fermions by $\psi = (\psi_1 - i \psi_2)/\sqrt{2}$ and $\psi^\dagger = (\psi_1  ^{\intercal}+ i \psi_2 ^{\intercal})/\sqrt{2}$. The matrix  $S$ can be written as
\begin{align}
S={}&m_{1,c}(\mathbb{I}_{2}\otimes G_{c}^{\text{S}})+m_{2,d}(\sigma_{2}\otimes G_{d}^{\text{A}})\notag\\
&+\Delta_{1,b}(\sigma_{3}\otimes S_{b})+\Delta_{2,b}(\sigma_{1}\otimes S_{b}), \label{eq:order_lagrangian}
\end{align}
and represents the $2N$-dimensional, symmetric, real-valued, traceless matrix. Here, $\Delta_{b}=\Delta_{1,b}-i\Delta_{2,b}$ are the superconducting, and $m_{1,c}$ and $m_{2,d}$ are the insulating order parameters \cite{so8}. They transform as symmetric tensor ($\text{dim}=N(N+1)/2$) and adjoint ($\text{dim}=N^{2}-1$) representations of SU($N$), respectively. $G_{c}^{\text{S}}$ ($c=1,\cdots,(N-1)(N+2)/2$) and $G_{d}^{\text{A}}$ ($d=1,\cdots, N(N-1)/2$) are the symmetric or antisymmetric generators of SU($N$), respectively \cite{so8}.

Eq.~\eqref{eq:HSM_lagrangian} is clearly invariant under the transformation $\phi\rightarrow (O\otimes\mathbb{I}_{2})\phi$ and $S\rightarrow OSO^{\intercal}$, where $O \in \text{SO}(2N)$. This means that taken together the $(N+1)(2N-1)$ Hubbard-Stratonovich real fields $(\Delta_{1,b},\Delta_{2,b},m_{a})$ form the symmetric irreducible second-rank tensor representation of \text{SO}($2N$) \cite{so8}.

\section{Field theory for the real symmetric tensor}\label{sec:tensor_field_theory}

Before proceeding further we should recall some of the features of the field theory for the \text{SO}($M$)-symmetric traceless tensor field. The Lagrangian with possible symmetry-allowed quartic terms is given by
\begin{align}
\mathcal{L}_{\text{sym}}={}&\frac{1}{\bar{M}_{\text{Tr}}}\left[\frac{1}{2}\text{Tr}[(\partial_{\tau} S)^{2}+(\nabla S)^{2}+rS^{2}]\right.\notag\\
&\quad\quad\quad\left.+\frac{\bar{\lambda}_{1}}{4\bar{M}_{\text{Tr}}}(\text{Tr}[S^{2}])^{2}+\frac{\bar{\lambda}_{2}}{4}\text{Tr}[S^{4}]\right],\label{eq:phi4}
\end{align}
where $S=\sum_{s=1}^{M_{s}}\varphi_{s}\mathbb{S}_{s}$, $M_{s}=(M-1)(M+2)/2$, and $\mathbb{S}_{a}$ are the linearly independent, real, symmetric, traceless $M$-dimensional matrices with $\text{Tr}[\mathbb{S}^{a}\mathbb{S}^{b}]=\bar{M}_{\text{Tr}}\delta_{ab}$. The norm of the matrices $\bar{M}_{\text{Tr}}$ is arbitrary, but we leave it unspecified, since its disappearance from the final results provides us with a useful check on the calculation. The Lagrangian is symmetric under $S\rightarrow O S O^ \top$, with $O\in \text{SO}(M)= \text{O}(M)/Z_2 $. The cubic term $\text{Tr}[S^3]$ is omitted since it can not emerge from the integration over fermions, which will be of our principal interest. $\mathcal{L}_{\text{sym}} $ then exhibits somewhat larger  $\text{SO}(M)\times Z_2 \simeq O(M)$ symmetry, where $S\rightarrow - S$ under the $Z_2$ transformation. This larger symmetry then guarantees that the cubic term is not generated under RG either.

For $M=2$ and $M=3$, $(\text{Tr}[S^{2}])^{2} = 2 \text{Tr}[S^{4}]$, so the quartic term can be written more simply as
\begin{align}
\frac{\bar{\lambda}_{1}}{4\bar{M}_{\text{Tr}}^{2}}&(\text{Tr}[S^{2}])^{2}+
\frac{\bar{\lambda}_{2}}{4\bar{M}_{\text{Tr}}}\text{Tr}[S^{4}]\notag\\
={}&\frac{1}{4}\left(\bar{\lambda}_{1}+\frac{\bar{M}_{\text{Tr}}}{M}\frac{M}{2}\bar{\lambda}_{2}\right)(\sum_{i=1}^{M_{s}}\varphi_{i}^{2})^{2},\label{eq:single_quartic}
\end{align}
and real coefficients $\varphi_{i}$ considered as belonging to the fundamental (vector) representation of \text{SO}(2) and \text{SO}(5), when $M=2,3$, respectively. For $M\geq4$, however, no such simplification is possible, and the ``single-trace" and ``double-trace" quartic invariants are independent. This is the relevant region for  the theory with fermions that we will be interested in, since there $M=2N$, and $N\geq 2$ because of the fermion doubling \cite{nielsen, herbuttimereversal}. Odd values of $M$, however, may also be relevant for strongly correlated electronic systems that exhibit fractionalization \cite{ray1, ray2, janssen2022}.

\subsubsection{Minimum of potential}\label{sec:conditions}

Depending on the signs of $\bar{\lambda}_{1,2}$ and the value of $M$, the Lagrangian in Eq.~\eqref{eq:phi4} has different minima \cite{delfino}.

\noindent1) When $\bar{\lambda}_{1}>0$ and $\bar{\lambda}_{2}<0$, at $r<0$ the minimum is at
\begin{align}
\bar{S}=\bar{\varphi}_{0}\left(\begin{matrix}
\mathbb{I}_{M-1}&0\\
0&-(M-1)\\
\end{matrix}\right),
\end{align}
where $\bar{\varphi}_{0}$ is the real-valued amplitude of the order parameter. At this minimum the $\text{SO}(M)$ symmetry is reduced to $\text{SO}(M-1)$. The amplitude $\bar{\varphi}_{0}$ of the order parameter is given by
\begin{align}
\bar{\varphi}_{0}={}&\pm\sqrt{\frac{-r}{(M/\bar{M}_{\text{Tr}})(M-1)\bar{\lambda}_{1}+(M^{2}-3M+3)\bar{\lambda}_{2}}},
\end{align}
and the interaction couplings $\bar{\lambda}_{1,2}$ should satisfy the stability condition
\begin{align}
 \bar{\lambda}_{1}+\frac{\bar{M}_{\text{Tr}}}{M}\frac{(M^{2}-3M+3)}{(M-1)}\bar{\lambda}_{2}>0.\label{eq:cond1}
\end{align}

\noindent2-1) When $\bar{\lambda}_{2}>0$, regardless of the sign of $\bar{\lambda}_{1}$, and for even $M$, the minimum configuration at $r<0$ is at,
\begin{align}
\bar{S}=\bar{\varphi}_{0}\left(\begin{matrix}
\mathbb{I}_{M/2}&0\\
0&-\mathbb{I}_{M/2}\\
\end{matrix}\right).
\end{align}
At this minimum the symmetry is broken down to  $\text{SO}(M/2)\times \text{SO}(M/2)$. The amplitude $m$ is given by
\begin{align}
\bar{\varphi}_{0}={}&\pm\sqrt{\frac{-r}{(M/\bar{M}_{\text{Tr}})\bar{\lambda}_{1}+\bar{\lambda}_{2}}},
\end{align}
and $\bar{\lambda}_{1,2}$ need satisfy the inequality
\begin{align}
\bar{\lambda}_{1}+\frac{\bar{M}_{\text{Tr}}}{M}\bar{\lambda}_{2}>0.\label{eq:cond2}
\end{align}

\noindent2-2) When $\bar{\lambda}_{2}>0$, again regardless of the sign of $\bar{\lambda}_{1}$, but now for odd $M$, at $r<0$ the minimum is at
\begin{align}
\bar{S}=\bar{\varphi}_{0}\left(\begin{matrix}
\mathbb{I}_{(M+1)/2}&0\\
0&-\frac{(M+1)}{(M-1)}\mathbb{I}_{(M-1)/2}\\
\end{matrix}\right),
\end{align}
so the symmetry is reduced to $\text{SO}((M+1)/2)\times \text{SO}((M-1)/2)$, and
\begin{align}
\bar{\varphi}_{0}={}&\pm\sqrt{\frac{-r}{\frac{M}{\bar{M}_{\text{Tr}}}\frac{(M+1)}{(M-1)}\bar{\lambda}_{1}+\frac{(M^{2}+3)}{(M-1)^{2}}\bar{\lambda}_{2}}}.
\end{align}
The stability condition now restricts $\bar{\lambda}_{1,2}$ to
\begin{align}
 \bar{\lambda}_{1}+\frac{\bar{M}_{\text{Tr}}}{M}\frac{(M^{2}+3)}{(M^{2}-1)}\bar{\lambda}_{2}>0.
\end{align}

If the couplings $\bar{\lambda}_{1,2}$ do not satisfy the appropriate inequality, the order parameter effective potential is not bounded from below and the stabilizing terms of the order higher than quartic need to be included. The symmetry breaking may then be expected to occur through a discontinuous (first-order) transition.

\subsubsection{RG flow to one-loop}\label{sec:boson_RG}

\begin{figure*}
\subfigure[]{
\includegraphics[width=0.18\textwidth]{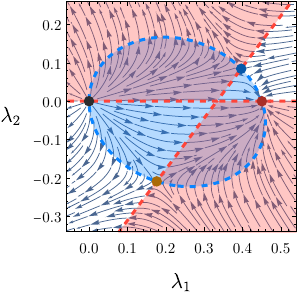}
\label{fig:order_RG1}}
\subfigure[]{
\includegraphics[width=0.18\textwidth]{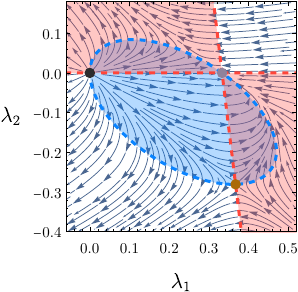}
\label{fig:order_RG2}}
\subfigure[]{
\includegraphics[width=0.18\textwidth]{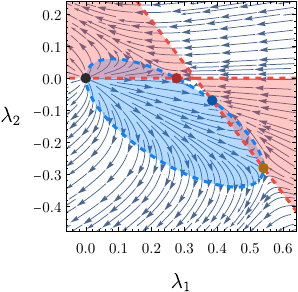}
\label{fig:order_RG3}}
\subfigure[]{
\includegraphics[width=0.18\textwidth]{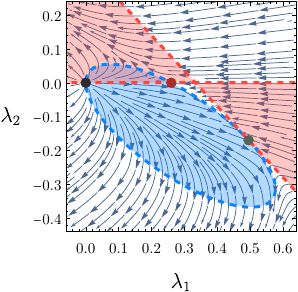}
\label{fig:order_RG4}}
\subfigure[]{
\includegraphics[width=0.18\textwidth]{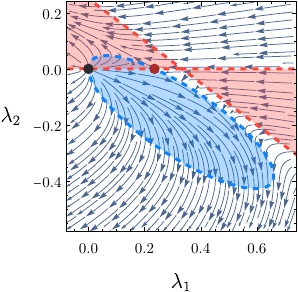}
\label{fig:order_RG5}}
\caption{The evolution of RG flow diagram for the field theory for the $\text{SO}(M)$-symmetric real tensor order parameter, without Yukawa coupling to fermions, as one varies the size of the matrix $M$: $M=3/2$, $M= M_{c1}\approx 2.702$, $M= 3.4$, $M= M_{c2}\approx3.624$, and $M= 4$, from left to right. The dark gray, red, blue, and orange points stand for the Gaussian, Wilson-Fisher, and two additional fixed points. The fixed points are located at the intersection of the blue and the red shaded areas, which represent the regions with positive $\partial\lambda_{1}/\partial\ell$ and positive $\partial\lambda_{2}/\partial\ell$, respectively. 
Here we set $\epsilon=1$.
}\label{fig:order_evolution}
\end{figure*}

One-loop beta-functions for the two quartic coupling constants near $3+1$ dimensions can also be computed:
\begin{widetext}
\begin{align}
\frac{d\lambda_{1}}{d\ell}={}&\epsilon\lambda_{1}-\frac{(M^{2}+M+14)}{8}\lambda_{1}^{2}-\frac{(2M^{2}+3M-6)}{4}\lambda_{1}\lambda_{2}-\frac{3(M^{2}+6)}{8}\lambda_{2}^{2},\label{eq:lambda1}\\
\frac{d\lambda_{2}}{d\ell}={}&\epsilon\lambda_{2}-3\lambda_{1}\lambda_{2}-\frac{(2M^{2}+9M-36)}{8}\lambda_{2}^{2},\label{eq:lambda2}
\end{align}
\end{widetext}
where $\epsilon=3-d$ and $\lambda_{1,2}$ are defined as
\begin{align}
\lambda_{1}={}&\frac{S_{d}\bar{\lambda}_{1}}{(2\pi)^{d}\Lambda^{d-3}},\quad
\lambda_{2}={}\frac{\bar{M}_{\text{Tr}}}{M}\frac{S_{d}\bar{\lambda}_{2}}{(2\pi)^{d}\Lambda^{d-3}}.
\end{align}

The dependence of the beta-functions on the arbitrary parameter $\bar{M}_{\text{Tr}}/M$ can be completely absorbed into the definition of $\lambda_{2}$, as expected. As a further check, since two quartic terms have the same form when $M=2,3$, the beta-functions in this case can be expressed as \cite{pelissetto}
\begin{align}
\frac{d}{d\ell}\tilde{\lambda}%
={}&\epsilon\tilde{\lambda}-\frac{(M^{2}+M+14)}{8}\tilde{\lambda}^{2},\label{eq:boson_RG23}
\end{align}
where
\begin{align}
\tilde{\lambda}={}&\lambda_{1}+\frac{M}{2}\lambda_{2}.
\end{align}
The beta-functions for $M=2,3$ reduce therefore to that of the textbook $\text{SO}(2)$ and $\text{SO}(5)$-symmetric vector $\phi ^4$-theories \cite{herbutbook}.

Let us analyze the RG flow equations for general $M$. First, the beta-functions always have the O($M_{s}$)-symmetric Wilson-Fisher fixed point at ${\lambda}_{2}=0$, and
\begin{align}
{\lambda}_{1}^{*}=\frac{8\epsilon}{M^{2}+M+14}.
\end{align}
The behavior of the RG flow in the $\lambda_{1}-\lambda_{2}$ plane crucially depends on $M$; in particular, there are two critical values of $M$,  $M_{c1}=(\sqrt{41}-1)/2 \approx 2.702$ and $M_{c2}\approx3.624$. When $M<M_{c1}$ (Fig.~\ref{fig:order_RG1}), the Wilson-Fisher fixed point is fully attractive, and in its domain of attraction governs the second-order phase transition. There also exist two fixed points of mixed stability, one with positive and the other with negative value of $\lambda_{2}$, and, of course,  the fully repulsive Gaussian fixed point, at $\lambda_1 = \lambda_2=0$. At $M=2$ the Wilson-Fisher and the mixed stability fixed point with $\lambda_2 >0$ lead to the same values of the critical exponents. In this sense the latter fixed point can be considered as  ``redundant" \cite{han}; it appears as an artifact of writing one and the same interaction term as two (seemingly) different terms.

As $M$ increases, the mixed-stability fixed point with positive $\lambda_{2}$ moves downward, so that at $M=M_{c1}$ it coincides with the Wilson-Fisher fixed point (Fig.~\ref{fig:order_RG2}). Upon further increase of $M$,  in the interval $M_{c1} < M < M_{c2}$, the two fixed points again become distinct and exchange stability: the Wilson-Fisher fixed point develops an unstable direction, and the other fixed point becomes fully stable. The stable fixed point then still resides in the region which allows the second-order phase transition (Fig.~\ref{fig:order_RG3}).  For $M=3$ in particular, we may again observe the same ``redundancy": the mixed-stable Wilson-Fisher fixed point and the fully stable fixed point are again equivalent and lead to the same critical exponents.

At $M=M_{c2}\approx3.624$ (Fig.~\ref{fig:order_RG4}), the stable fixed point coincides with another fixed point with negative $\lambda_{2}$.
For $M > M_{c2}$ (Fig.~\ref{fig:order_RG5}), the two fixed points split and become complex, so in the real $\lambda_1 - \lambda_2$ plane there are only the  Gaussian and the Wilson-Fisher fixed point of mixed stability left. For $M > M_{c2}$, which includes the first value that will be of relevance here of $M=4$ there is therefore no fully stable fixed point in the real $\lambda_1 - \lambda_2$ plane. In particular, starting the flow at any positive $\lambda_2$, the value of $\lambda_1$ eventually turns negative, and the flow ends up in the unstable region where the phase transition is expected to become fluctuation-induced first order. Higher-order beta-functions both in $\epsilon$-expansion and fixed-dimension RG fail to find a stable fixed point. In particular, the value of $M_{c2}$ was found to vary very little as $\epsilon$ varies between zero and one, as well as to be only weakly dependent on the scheme of calculation \cite{pelissetto}. Nevertheless, a Monte Carlo simulation on an intimately related lattice model \cite{pelissetto} suggests a possible continuous transition for $M=4$. Conformal bootstrap study \cite{reehorst} also finds features that seem consistent with a continuous transition at $M=4$. The issue is still open at the present time. The transition at $M=6$, however, appears to be first order \cite{bonati}.

It is interesting that, in contrast to what happens in the standard $\phi ^4$-theories for vector order parameters, in  the large-$M$ limit there is no stable fixed point in the $\lambda_{1}-\lambda_{2}$ plane. One may wonder why this is, considering that the term proportional to $\lambda_1 \lambda_2$ in the beta-function for $\lambda_2$ in this limit becomes negligible, and the $\lambda_2$ beta-function shows an attractive fixed point at $\lambda_2 ^* = 4\epsilon/M^2$. The  reason is that the  RG flow with $\lambda_2$ alone  is not closed, and a finite $\lambda_2$ {\it generates} $\lambda_1$ even if $\lambda_1$ was initially absent. Once  $\lambda_1$ is generated, all the terms in its own beta-function at large $M$ are of the same order, and the beta-function for $\lambda_1$ ends up being negative-definite. Its zeros in the large-$M$ limit are found at $\lambda_1 ^* = -4\epsilon(1\pm i \sqrt{2})/M^2$, and therefore are complex even at large $M$.

Recalling that by our definitions $\lambda_1 \sim \bar{\lambda}_1$ and $\lambda_2 \sim \bar{\lambda}_2 /M$, one could say that in terms of the original couplings the fixed point is at $\bar{\lambda}_2 \sim 1/M$ and real, and $\bar{\lambda}_1 \sim 1/M^2$ and complex. In this rather restricted sense, if one neglects the generation of the double-trace quartic term which has a coupling that becomes $O(1/M^2)$ at the fixed point, to the leading order in small $1/M$ there actually is a stable fixed point at large $M$. It is destabilized, however, by the the next-order $\sim 1/M^2$ terms, which inevitably appear during the RG process.

To summarize, according to the one-loop (and higher-loop) beta-functions the second-order phase transition occurs only when $M < M_{c2}\approx 3.6$, otherwise, the transition is weakly first-order for any coupling, and the RG leads to the runaway flow. This is the result in the pure bosonic field theory for the symmetric traceless real matrix order parameter, without any coupling to fermions. We will see next that the conclusion changes if we add the Yukawa interaction to a sufficient number of fermion flavors. The evolution of the RG flow diagram  in $\lambda_{1}$-$\lambda_{2}$ plane with $M$ is presented in Fig.~\ref{fig:order_evolution}

\section{GNY theory with \text{SO}($2N$) symmetric tensor fields}\label{sec:GNY}

Let us consider the Gross-Neveu-Yukawa theory for the Majorana representation with \text{SO}($2N$) symmetric tensor order parameters. The action is given by
\begin{widetext}
\begin{align}
\bar{\mathcal{S}}_{\text{GNY}}={}&\int d\tau d^{d}x\left[\frac{1}{2}(\phi^{\intercal}(\mathbb{I}_{M}\otimes(\mathbb{I}_{2}\partial_{\tau}-iv_{F}\sigma_{1}\partial_{1}-iv_{F}\sigma_{3}\partial_{2}))\phi)+\frac{\bar{g}}{2}(\phi^{\intercal}(S\otimes\sigma_{2})\phi)
\right]\notag\\{}
&+\int d\tau d^{d}x
\frac{1}{\bar{M}_{\text{Tr}}}\left[\frac{1}{2}\text{Tr}[(\partial_{\tau} S)^{2}+v_{B}^{2}(\nabla S)^{2}+rS^{2}]+\frac{\bar{\lambda}_{1}}{4\bar{M}_{\text{Tr}}}(\text{Tr}[S^{2}])^{2}+\frac{\bar{\lambda}_{2}}{4}\text{Tr}[S^{4}]\right],
\end{align}
\end{widetext}
with the real, traceless, $M$-dimensional symmetric matrix $S$ as already defined, and $M=2N$. Using this action we then compute the RG equation at the one-loop order, which read as follows:
\begin{widetext}
\begin{align}
\frac{dy}{d\ell}={}&\alpha_{g}\left[\frac{M_{s}y^{2}}{3(1+y)^{2}}+\frac{M(1+y)}{16}\right]y(1-y),\\
\frac{d\alpha_{g}}{d\ell}%
={}&\epsilon \alpha_{g}-\alpha_{g}^{2}\frac{M_{s}y^{2}}{(1+y)^{2}}(1-y)-\alpha_{g}^{2}[\frac{M_{s}y^{3}}{(1+y)^{2}}+\frac{M}{8}+\frac{(M-2)y^{2}}{2(1+y)}],\\
\frac{d\lambda_{1}}{d\ell}%
={}&\epsilon \lambda_{1}-\frac{M\alpha_{g}\lambda_{1}}{4}+\frac{3M(1-y^{2})}{16}\alpha_{g}\lambda_{1}-\frac{(M^{2}+M+14)}{8}\lambda_{1}^{2}-\frac{(2M^{2}+3M-6)}{4}\lambda_{1}\lambda_{2}-\frac{3(M^{2}+6)}{8}\lambda_{2}^{2},\\
\frac{d\lambda_{2}}{d\ell}%
={}&\epsilon \lambda_{2}-\frac{M\alpha_{g}\lambda_{2}}{4}+\frac{3M(1-y^{2})}{16}\alpha_{g}\lambda_{2}-3\lambda_{1}\lambda_{2}-\frac{(2M^{2}+9M-36)}{8}\lambda_{2}^{2}+\alpha_{g}^{2}\frac{My^{3}}{4}
\end{align}
\end{widetext}
where
\begin{align}
y={}&\frac{v_{F}}{v_{B}},\quad\quad \alpha_{g}=\frac{\bar{M}_{\text{Tr}}}{M}\frac{\bar{g}^{2}S_{d}}{(2\pi)^{d}v_{F}^{d}\Lambda^{d-3}},\\
 \lambda_{1}={}&\frac{\bar{\lambda}_{1}S_{d}}{(2\pi)^{d}v_{B}^{d}\Lambda^{d-3}},\quad \lambda_{2}=\frac{\bar{M}_{\text{Tr}}}{M}\frac{\bar{\lambda}_{2}S_{d}}{(2\pi)^{d}v_{B}^{d}\Lambda^{d-3}}.\label{eq:lambda_def}
\end{align}
Dependence on the arbitrary normalization $\bar{M}_{\text{Tr}}/M$ again disappears once it is absorbed into the definition of $\alpha_{g}$ and $\lambda_{2}$.

Note that if one starts the flow with $\lambda_1 = \lambda_2 =0$, integration over fermions generates directly only the single-trace quartic term, i.\,e.~the self-interaction $\lambda_2$, which, once generated, itself generates the double-trace self-interaction interaction coupling $\lambda_1$.

The parameter $y$ has a stable fixed point value at unity, so there is again the emergent Lorentz symmetry. Hereafter we therefore set $y=1$.
If we turn off the Yukawa coupling $\alpha_{g}$, the remaining equations reduce to the previously discussed RG flow equations for $\lambda_{1,2}$.
When $\alpha_{g}\neq0$, $\alpha_{g}$ flows under RG into a  finite stable fixed-point value,
\begin{align}
\alpha_{g}^{*}=\frac{8\epsilon }{(M^{2}+4M-6)}.
\end{align}
Inserting $\alpha_{g}^{*}$ into the remaining flow equations of $\lambda_{1,2}$ one finds
\begin{widetext}
\begin{align}
\frac{d\lambda_{1}}{d\ell}
={}&\frac{(M^{2}+2M-6)}{(M^{2}+4M-6)}\epsilon \lambda_{1}-\frac{(M^{2}+M+14)}{8}\lambda_{1}^{2}-\frac{(2M^{2}+3M-6)}{4}\lambda_{1}\lambda_{2}-\frac{3(M^{2}+6)}{8}\lambda_{2}^{2},\\
\frac{d\lambda_{2}}{d\ell}
={}&\frac{(M^{2}+2M-6)}{(M^{2}+4M-6)}\epsilon \lambda_{2}-3\lambda_{1}\lambda_{2}+\frac{(2M^{2}+9M-36)}{8}\lambda_{2}^{2}+\frac{16M\epsilon^{2}}{(M^{2}+4M-6)^{2}}.\label{eq:lambda2_with_fermion}
\end{align}
\end{widetext}
These flow equations again do not have a stable fixed point for any $M\geq4$. This is easiest to see in the large-$M$ limit: since the number of the bosonic components of the order parameter grows as $M_s\sim M^2$ and the number of fermionic components only as $M$, in the large-$M$ limit the bosonic contribution to the beta-function for the Yukawa coupling dominates, and the fixed point $\alpha_{g}^{*} \sim 1/M^2$. This way Yukawa coupling drops out of the remaining two beta-functions in the large-$M$ limit, and one is left with the pure bosonic theory which we already saw leads only to runaway flow and the first order transition. The conclusion then remains the same for all values of $M$.

\section{GNY theory with $\text{SO}(2N)$ symmetric tensor fields and $N_{f} >1$ fermion flavors}\label{sec:GNY_Nf}

\begin{figure*}
\subfigure[]{
\includegraphics[width=0.18\textwidth]{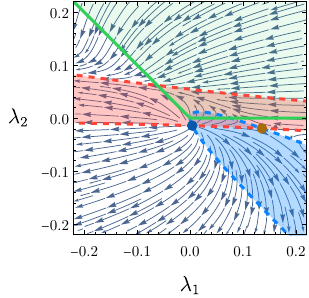}
\label{fig:order_RG1_fermion}}
\subfigure[]{
\includegraphics[width=0.18\textwidth]{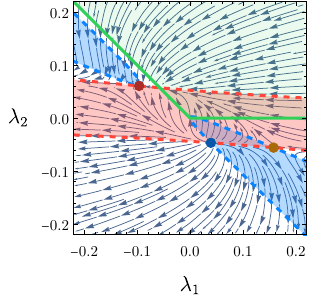}
\label{fig:order_RG2_fermion}}
\subfigure[]{
\includegraphics[width=0.18\textwidth]{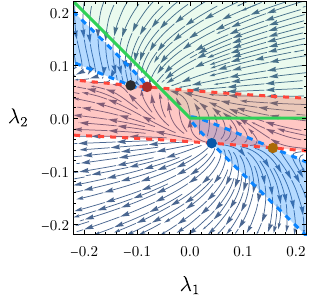}
\label{fig:order_RG3_fermion}}
\subfigure[]{
\includegraphics[width=0.18\textwidth]{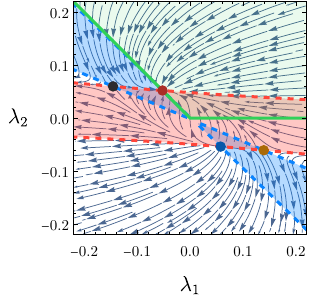}
\label{fig:order_RG4_fermion}}
\subfigure[]{
\includegraphics[width=0.18\textwidth]{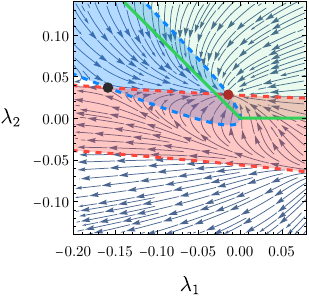}
\label{fig:order_RG5_fermion}}
\caption{The evolution of the RG flow diagrams for fixed $N=4$ ($M=8$) with the number of Majorana fermions $N_{f}$: $N_{f}=1$, $N_{f}= N_{f,c1}\approx 7.18$, $N_{f}=7.5$, $N_{f}=N_{f,c2}\approx10.80$, and $N_{f}=40$, from left to right. The dots stand for fixed points, and the blue and red shaded area again represent the regions with positive $\partial\lambda_{1}/\partial\ell$ and positive $\partial\lambda_{2}/\partial\ell$, respectively. The green shaded area (in between the two green lines) stands for the stability condition Eq.~\eqref{eq:cond1}, and yields a continuous transition with the symmetry breaking pattern  $\text{SO}(M)\rightarrow\text{SO}(M/2)\otimes\text{SO}(M/2)$. 
Here we set $\epsilon=1$.
}\label{fig:order_evolution_fermion}
\end{figure*}

To enable the Yukawa coupling to qualitatively modify the flow of the self-interaction couplings $\lambda_1$ and $\lambda_2$ we introduce $N_f >1$ flavor of Majorana fermions, each one still with $4N$ components. This way the theory acquires an additional flavor symmetry SO($N_{f}$), under which the bosonic symmetric tensor field transforms as a scalar. The action is then given by
\begin{widetext}
\begin{align}
\mathcal{S}={}&\int d\tau d^{d}x\left[\frac{1}{2}(\phi^{\intercal}(\mathbb{I}_{N_{f}}\otimes\mathbb{I}_{M}\otimes(\mathbb{I}_{2}\partial_{\tau}-iv_{F}\sigma_{1}\partial_{1}-iv_{F}\sigma_{3}\partial_{2}))\phi)+\frac{\bar{g}}{2}(\phi^{\intercal}(\mathbb{I}_{N_{f}}\otimes S\otimes\sigma_{2})\phi)\right]\notag\\
{}&+\int d\tau d^{d}x\;\frac{1}{\bar{M}_{\text{Tr}}}\left[\frac{1}{2}\text{Tr}[(\partial_{\tau} S)^{2}+v_{B}^{2}(\nabla S)^{2}+r S^{2}]+\frac{\bar{\lambda}_{1}}{4\bar{M}_{\text{Tr}}}(\text{Tr}[S^{2}])^{2}+\frac{\bar{\lambda}_{2}}{4}\text{Tr}[S^{4}]\right],
\end{align}
with the Majorana field $\phi$ now having $2 M N_f$ components.

The RG flow equations at one loop  are now
\begin{align}
\frac{dy}{d\ell}={}&\alpha_{g}[\frac{M_{s}y^{2}}{3(1+y)^{2}}+\frac{MN_{f}(1+y)}{16}]y(1-y),\\
\frac{d\alpha_{g}}{d\ell}%
={}&\epsilon \alpha_{g}-\alpha_{g}^{2}\frac{M_{s}y^{2}}{(1+y)^{2}}(1-y)-\alpha_{g}^{2}[\frac{M_{s}y^{3}}{(1+y)^{2}}+\frac{MN_{f}}{8}+\frac{(M-2)y^{2}}{2(1+y)}],\\
\frac{d\lambda_{1}}{d\ell}%
={}&\epsilon \lambda_{1}-\frac{MN_{f}\alpha_{g}\lambda_{1}}{4}+\frac{3MN_{f}(1-y^{2})}{16}\alpha_{g}\lambda_{1}-\frac{(M^{2}+M+14)}{8}\lambda_{1}^{2}-\frac{(2M^{2}+3M-6)}{4}\lambda_{1}\lambda_{2}-\frac{3(M^{2}+6)}{8}\lambda_{2}^{2},\\
\frac{d\lambda_{2}}{d\ell}%
={}&\epsilon \lambda_{2}-\frac{MN_{f}\alpha_{g}\lambda_{2}}{4}+\frac{3MN_{f}(1-y^{2})}{16}\alpha_{g}\lambda_{2}-3\lambda_{1}\lambda_{2}-\frac{(2M^{2}+9M-36)}{8}\lambda_{2}^{2}+\alpha_{g}^{2}\frac{MN_{f}y^{3}}{4},
\end{align}
with $y$, $\alpha_{g}$, and $\lambda_{1,2}$ defined the same as before. The beta-functions reduce to the previously derived set when $N_{f}\rightarrow1$. 
Note that one can get the beta functions for $MN_{f}$ copies of two-component Weyl fermions instead of the Majorana fermions in two dimensions, or $MN_{f}$ copies of four-component Majorana fermions in three dimensions by replacing $N_{f}\rightarrow 2N_{f}$.

At the stable Lorentz-symmetric fixed point $y^{*}=1$, the remaining three flow equations become
\begin{align}
\frac{d\alpha_{g}}{d\ell}
={}&\epsilon \alpha_{g}-\frac{\alpha_{g}^{2}}{8}[2M_{s}+MN_{f}+2(M-2)],\\
\frac{d\lambda_{1}}{d\ell}
={}&\epsilon \lambda_{1}-\frac{MN_{f}\alpha_{g}\lambda_{1}}{4}-\frac{(M^{2}+M+14)}{8}\lambda_{1}^{2}-\frac{(2M^{2}+3M-6)}{4}\lambda_{1}\lambda_{2}-\frac{3(M^{2}+6)}{8}\lambda_{2}^{2},\\
\frac{d\lambda_{2}}{d\ell}
={}&\epsilon \lambda_{2}-\frac{MN_{f}\alpha_{g}\lambda_{2}}{4}-3\lambda_{1}\lambda_{2}-\frac{(2M^{2}+9M-36)}{8}\lambda_{2}^{2}+\alpha_{g}^{2}\frac{MN_{f}}{4},
\end{align}
and the first one has the stable fixed point at the Yukawa coupling
\begin{align}
\alpha_{g}^{*}={}&\frac{8\epsilon }{(M^{2}+(3+N_{f})M-6)}.
\end{align}
For $M \gg N_f$, $\alpha_{g}^{*} \sim 1/M^2$, as before. In the opposite limit $N_f \gg M$, on the other hand, $\alpha_{g}^{*} = 8\epsilon / ( M N_f)$. In this limit we will finally discover a stable fixed point of the RG.

Inserting $\alpha_{g}^{*}$ into the remaining flow equations of $\lambda_{1,2}$ we finally obtain
\begin{align}
\frac{d\lambda_{1}}{d\ell}
={}&\frac{(M^{2}+M(3-N_{f})-6)}{(M^{2}+M(3+N_{f})-6)}\epsilon \lambda_{1}-\frac{(M^{2}+M+14)}{8}\lambda_{1}^{2}-\frac{(2M^{2}+3M-6)}{4}\lambda_{1}\lambda_{2}-\frac{3(M^{2}+6)}{8}\lambda_{2}^{2},\label{eq:lambda1_fixed}\\
\frac{d\lambda_{2}}{d\ell}
={}&\frac{(M^{2}+M(3-N_{f})-6)}{(M^{2}+M(3+N_{f})-6)}\epsilon \lambda_{2}-3\lambda_{1}\lambda_{2}-\frac{(2M^{2}+9M-36)}{8}\lambda_{2}^{2}+\frac{16MN_{f}\epsilon^{2}}{(M^{2}+(N_{f}+3)M-6)^{2}}.\label{eq:lambda2_fixed}
\end{align}
\end{widetext}

It now becomes evident how a stable fixed point emerges in the large-$N_f$ (fixed-$N$) limit: (1) large number of fermions yields the boson anomalous dimension close to $\epsilon$, so the linear terms in both beta-functions become $(\epsilon- 2 \eta_\phi)\lambda_{1,2} \rightarrow - \epsilon \lambda_{1,2}$, (2) the last term in the beta-function Eq.~\eqref{eq:lambda2_fixed} becomes $16 \epsilon^2 / (M N_f)$, so balancing it against the first (linear term) and neglecting the other two terms yields $\lambda_2 ^* = 16 \epsilon/(M N_f)$, and (3) inserting this fixed point value into the remaining beta-function Eq.~\eqref{eq:lambda1_fixed} finally gives $\lambda_1 ^* = -96\epsilon/N_f ^2$ and  real. In the large-$N_f$ limit only the first and the last term in both beta-functions matter, and $|\lambda_1 ^* |  \ll \lambda_2 ^* $, which then justifies the above simplified calculation of the fixed points. The hierarchy between the fixed point values of the couplings that emerges in the large-$N_f$ limit is a  consequence of the fact that the integration over fermions directly generates only $\lambda_2$, as we have noticed earlier.

\begin{figure}
\centering
\includegraphics[width=\linewidth]{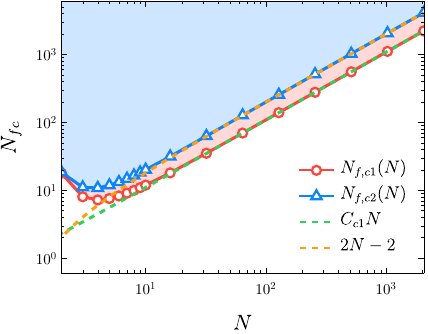}
\caption{The red circle and blue triangle markers are the values of $N_{f,c1}$ and $N_{f,c2}$ at various values of $N$. The green and orange dashed lines are fitting functions of $N_{f,c1}$ and $N_{f,c2}$ in large $N$ limit. Below the red circle marker, the RG flow runs away. Between the blue triangle and red circle markers (red shaded region), the beta functions have the real stable fixed point in the unstable region, so the transition should be first-order. Only above the blue triangle markers (blue shaded region) do the beta-functions possess the stable fixed point that satisfies the condition for the second-order phase transition. 
}\label{fig:critical_Nf}
\end{figure}

\subsection{Critical $N_{f}$ for continuous phase transition}

When $N_{f}=1$, there is no stable fixed point on the $\lambda_{1}$-$\lambda_{2}$ plane for any $M\geq4$, even with the Yukawa coupling at the non-trivial fixed point value $\alpha_{g} = \alpha_{g}^{*}$. However, at large $N_f$  we found that the stable fixed point with $\lambda_2 ^* \gg |\lambda_1 ^*| $ actually exists. By continuity there exists then a critical value of $N_f$ where the stable fixed point becomes real.

One can discern, however, not one but two critical values: $N_{f,c1}$ and $N_{f,c2}$. When $N_{f}<N_{f,c1}$, there is no stable fixed point (Fig.~\ref{fig:order_RG1_fermion}), and at  $N_{f}= N_{f,c1}$, the stable fixed point first emerges from the complex plane and becomes real (Fig.~\ref{fig:order_RG2_fermion}). At large $M$, $N_{f,c1}$ is given by $N_{f,c1}\approx  C_{c1}N$  where the constant can be analytically computed to be $C_{c1}\approx 2(3^{3/2}-4-\sqrt{6(7-4\sqrt{3})}\approx1.080$. The fixed point right at $N_{f}=N_{f,c1}$ in large $M$ is given by
\begin{align}
(\lambda_{1}^{*},\lambda_{2}^{*})=(-\frac{\sqrt{2+\sqrt{3}}}{N^{2}}\epsilon ,\frac{\sqrt{2+\sqrt{3}}}{\sqrt{3}N^{2}}\epsilon ).
\end{align}
However, the existence of the stable fixed point alone does not suffice for the continuous phase transition in the GNY theory; the fixed point values also need to satisfy the stability conditions as discussed in Sec.~\ref{sec:conditions}. The relevant condition is Eq.~\eqref{eq:cond2}, which implies that the couplings at fixed point need to satisfy
\begin{equation}
\lambda_1 ^* + \lambda_2 ^* >0
\end{equation}
to correspond to the second-order phase transition. The condition, however, is violated by the values of the fixed points right at $N_{f}= N_{f,c1}$ at all $N\geq2$ ($M\geq4$). At large-$N_f$, on the other hand, we found $\lambda_1 ^* \gg |\lambda_2 ^*|$, and the fixed point clearly lies in the stable region. So
by increasing $N_{f}$ further one must detect the second critical value $N_{f,c2}$ (Fig.~\ref{fig:order_RG4_fermion}), so that for $N_{f}>N_{f,c2}$, the fixed point values satisfies the condition for the second-order phase transition (Fig.~\ref{fig:order_RG5_fermion}). In the large-$M$ limit in particular, $N_{f,c2}\approx 2N-2$, and the fixed point values right at $N_{f}=N_{f,c2}$ is given by
\begin{align}
(\lambda_{1}^{*},\lambda_{2}^{*})%
\approx(-\frac{\epsilon }{N^{2}},\frac{\epsilon }{N^{2}})+\mathcal{O}(N^{-3}).%
\end{align}
Since the slope of $N_{f,c2}$ vs.~$N$ is larger than that of $N_{f,c1}$, $N_{f,c1} < N_{f,c2}$ (Fig.~\ref{fig:critical_Nf}).

Note also that the stable fixed point is always with the coupling $\lambda_{2}>0$, which implies that the symmetry breaking pattern is $\text{SO}(2N)\rightarrow\text{SO}(N)\times\text{SO}(N)$. This is consistent with the previous mean-field analysis of Ref.~\cite{so8}.

\begin{figure}
    \centering
    \includegraphics[width=\linewidth]{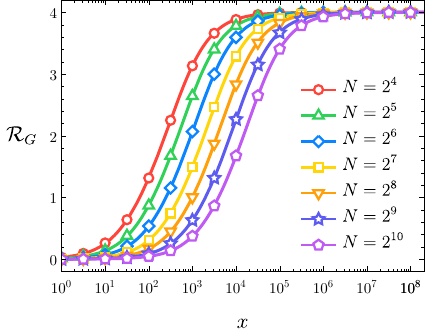}
    \caption{The mass gap ratio $\mathcal{R}_{G}$ for $N=2^{4},\cdots,2^{10}$ and varying $x=N_{f}-N_{f,c2}(N)$. At $x=0$ ($N_{f}=N_{f,c2}$), $\mathcal{R}_{G}(N,x=0)=0$, and for large $x$ limit, $\mathcal{R}_{G}(N,x\rightarrow\infty)\rightarrow4$ regardless of the value of $N$. The markers are actual numerical values of $\mathcal{R}_{G}(N,x)$ and the solid lines are the results from the fitting functions $f_{G}(N,x)$, respectively.
    }
    \label{fig:mass_gap_ratio}
\end{figure}

\subsection{Mass gap ratio}
Let us consider the mass gap ratio at $N_{f}=N_{f,c2}$. Since the symmetry is broken as $\text{SO}(2N)\rightarrow\text{SO}(N)\times\text{SO}(N)$, the mass ratio is given by
\begin{align}
\mathcal{R}_{G}=\frac{\bar{M}_{\text{Tr}}}{M}\frac{m_{\varphi}^{2}}{m_{\psi}^{2}}=\frac{2(\lambda_{1}^{*}+\lambda_{2}^{*})}{\alpha_{g}^{*}},
\end{align}
where we used the definitions in Eq.~\eqref{eq:lambda_def}. As discussed above, the ratio vanishes right at $N_{f}=N_{f,c2}$, since this critical value is defined by the saturation of the inequality, and $\lambda_{1}^{*}=-\lambda_{2}^{*}$. Numerical computation yields that $\mathcal{R}_{G} \rightarrow 4$ in the large $N_{f}$ limit for fixed $N$, similarly as in to Eq.~\eqref{eq:massgap_GN} for the Ising GNY model in the limit $N\rightarrow \infty$ (see Fig.~\ref{fig:mass_gap_ratio}). Let $x=N_{f}-N_{f,c2}(N)$. The mass gap ratio is then well-fitted by the following function
\begin{align}
    f_{G}(N,x)=\frac{a_{0}(N)+a_{1}(N)x+4x^{2}}{b_{0}(N)+a_{1}(N)x+x^{2}},\label{eq:fitting}
\end{align}
where the coefficients $a_{0}$ and $b_{0}$, and $a_{1}$ and $b_{1}$ are proportional to $N^{2}$ and $N$, respectively.

\subsection{Correlation length exponent}

\begin{figure}
    \centering
    \includegraphics[width=\linewidth]{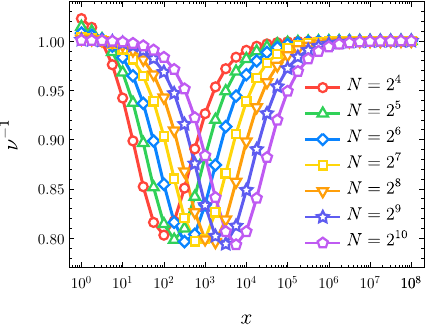}
    \caption{The correlation length exponent $\nu^{-1}$ in terms of $x=N_{f}-N_{f,c2}(N)$ for given $N$ when $\epsilon=1$. For large $x$, it converges to $1$.}
    \label{fig:correlnu}
\end{figure}

To obtain the correlation length exponent we compute the usual mass renormalization of the order parameter. To one-loop then
\begin{align}
\nu^{-1}={}&2-\frac{1}{8}[(M^{2}+M+2)\lambda_{1} ^{*}+(2M^{2}+3M-6)\lambda_{2}^{*} ]\notag\\&-\frac{MN_{f}}{8}\alpha_{g}.
\end{align}
 In the large-$M$ limit, at $N_{f}=N_{f,c2}\approx 2N-2$, for example, the correlation length exponent is given by
\begin{align}
\nu^{-1}={}&2-\frac{(M^{2}+2M-8)}{2M^{2}}\epsilon -\frac{M(M-2)}{(2M^{2}+M-6)}\epsilon \notag\\
\approx{}&2-\epsilon +\mathcal{O}(M^{-1}).
\end{align}
The correlation length exponent for $N_{f}>N_{f,c2}$ is presented in Fig.~\ref{fig:correlnu}.

\subsection{Anomalous dimensions}

The anomalous dimensions of fermion and boson are similarly given by one-loop expressions
\begin{align}
\eta_{\psi}={}&\alpha_{g}^{*}\frac{M_{s}}{8}
={} \frac{M_{s}\epsilon }{(M^{2}+(3+N_{f})M-6)},\\
\eta_{\varphi}={}&\alpha_{g}^{*}\frac{MN_{f}}{8}
={}\frac{MN_{f}\epsilon }{(M^{2}+(3+N_{f})M-6)}.
\end{align}
Note that in the large-$N_f$ limit, fermion's anomalous dimension $\eta_{\psi} \rightarrow (M \epsilon) /(2 N_f) \rightarrow 0$, whereas $\eta_{\varphi} \rightarrow \epsilon \rightarrow 1$. The latter result reflects the fact that bosonic field in this limit acquires its dynamics entirely from fermions, and consequently the bosonic propagator (at critical point) becomes proportional to $1/Q^{2-\eta_\varphi}\sim 1/Q$, in $2+1$ dimensions.

\subsection{Size of first-order phase transition}

We saw that for $N_{f} < N_{f,c2}$, the RG flow either runs away or it has a stable fixed point that does not satisfy the stability condition, so in either case the phase transition should be first-order. Let us define and discuss the ``size" of the first-order phase transition \cite{herbutbook}, and in particular consider its dependence on $N$.

We begin with the order parameter effective potential and add the simplest positive next-order (sextic) term to stabilize it:
\begin{align}
V(\Phi)={}&\frac{r}{2\bar{M}_{\text{Tr}}}\text{Tr}[\Phi^{2}]+\frac{\bar{\lambda}_{1}}{4\bar{M}_{\text{Tr}}^{2}}(\text{Tr}[\Phi^{2}])^{2}\notag\\
&+\frac{\bar{\lambda}_{2}}{4\bar{M}_{\text{Tr}}}\text{Tr}[\Phi^{4}]+\frac{\bar{\kappa}}{6\bar{M}_{\text{Tr}}}\text{Tr}[\Phi^{6}].
\end{align}
Assuming the broken symmetry solution $\Phi_{0}$ at the non-trivial minimum has the form,
\begin{align}
\Phi_{0}=\bar{\varphi}_{0}\left(\begin{matrix}
\mathbb{I}_{N}&0\\
0&-\mathbb{I}_{N}\\
\end{matrix}\right),
\end{align}
so that the symmetry is broken as $\text{SO}(2N)\rightarrow \text{SO}(N)\times \text{SO}(N)$, the amplitude $\bar{\varphi}_{0}$ at the first-order transition can be taken to be one measure of its ``size". At the point of discontinuous transition $V(0)=V(\Phi_{0})$, we can find the finite value of the tuning parameter $r$ at the first-order phase transition to be,
\begin{align}
r_{c}=\frac{3(\bar{\lambda}_{1}+(\bar{M}_{\text{Tr}}/M)\bar{\lambda}_{2})^{2}}{16(\bar{M}_{\text{Tr}}/M)^{2}\bar{\kappa}},
\end{align}
where $\bar{\lambda}_{1}+(\bar{M}_{\text{Tr}}/M)\bar{\lambda}_{2}<0$ for the first-order phase transition. Then, the amplitude of the order parameter at the point of transition is
\begin{align}
\bar{\varphi}_{0}^2 =\left(\frac{3r_{c}}{\bar{\kappa}}\right)^{1/2}=\left(\frac{3|\bar{\lambda}_{1}+(\bar{M}_{\text{Tr}}/M)\bar{\lambda}_{2}|}{4(\bar{M}_{\text{Tr}}/M)\bar{\kappa}}\right),
\end{align}
and proportional to to $r_{c}^{1/2}$. One can therefore also use the finite value of the tuning parameter at the point of transition $r_{c}$ as a measure of its size.

Furthermore, we can connect $r_c$ to the RG time $l=\ln b$ it takes to reach the boundary of the stability region as $r_c \sim b^{-2}$ \cite{herbutbook}. Clearly this is a non-universal quantity since it depends on the initial condition. Here we limit ourselves to the values of $N_f/M \ll 1$, where the consideration simplifies. In this regime the Yukawa fixed point is at $\alpha_g ^* = 8 \epsilon/M^2$, and the term proportional to $\lambda_1 \lambda_2$ in the beta-function for $\lambda_2$ becomes negligible. The fixed point for $\lambda_2$ is therefore at
\begin{equation}
\lambda_2 ^* = \frac{4 \epsilon}{M^2} \left( 1 + 2 (N_{f}/M)  - 8 (N_{f}/M)^{2} + \mathcal{O}((N_{f}/M)^{3}) \right).
\end{equation}
We can therefore simply insert the fixed point values $\alpha_g ^*$ and $\lambda_2 ^*$ into the beta function for $\lambda_1$ and integrate. Since the stability boundary is at $\lambda_1 + \lambda_2=0$, if the flow starts from $\lambda_1=0$ and $\lambda_2 = \lambda_2 ^*$,  $\lambda_1$ near the boundary becomes also $\sim 1/M^2$ and its neglect in the beta-function for $\lambda_2$ is justified. Integrating, one finds the RG time along the trajectory of fixed $\alpha_g = \alpha_g ^*$  and $\lambda_2 = \lambda_2 ^*$ from $\lambda_1=0$ to $\lambda_1 = -\lambda_2 ^* $ to be
\begin{equation}
l= 0.87 + 4.70 (N_{f}/M)^{2}  + \mathcal{O}((N_{f}/M)^{3}).
\end{equation}
The RG time, for this initial condition at least, to reach the unstable region approaches a constant and decreases with $M$, at large $M$. The size of the first-order transition in this regime should therefore increase with $M$.

\section{Conclusion and discussion}\label{sec:discussion}

To summarize, we constructed and analyzed the relativistic GNY field theory that features  the symmetric second-rank real irreducible $\text{SO}(2N)$ tensor Yukawa-coupled to $N_f$ flavors of $4N$-component Majorana fermions. The field theory describes the spontaneous breaking of the $\text{SO}(2N)$ flavor symmetry of the GN model in 2+1 dimensions to $\text{SO}(N)\times \text{SO}(N)$. The Lagrangian contains two self-interaction quartic terms, and shows only runaway RG flows for a range of small $N_f$, which  includes the value of $N_f=1$ relevant to the GN model. We interpret this as the sign of the transition being fluctuation-induced first-order. This is then in accord with the conclusion of our earlier study of the same transition directly on the GN model \cite{han}. Above the critical value of fermion flavors, however, a real stable fixed point of the RG flow emerges within the expansion near the upper critical dimension of 3+1. The critical value of $4N$-component Majorana fermions $N_f \approx 2N$ for $N\gg 1$, and has a minimum near $N\approx 4$.

The mean-field solution in the previous study \cite{so8} is now seen to correspond to the large-$N_f$ limit of fermion flavors taken at fixed number of components of each fermion. In this limit the fixed point is at the Yukawa coupling $\alpha_{g}^{*}$ and the single-trace self-interaction $\lambda_{2}^* $ which are both $\mathcal{O}(1/N_f)$, and at the  double-trace self-interaction $\lambda_{1}^{*}= \mathcal{O}(1/N_f^2) $. Since $\lambda_{1}$ becomes negligible, the fixed point values satisfy the stability condition, and imply a continuous phase transition.
The identification is also supported by the fact that the mean-field solution corresponds to the minimum found for $\lambda_{2}>0$, as discussed in Sec.~\ref{sec:conditions}. Integrating over fermions generates the potential for the matrix order parameter with only the single-trace $\text{Tr}[S^{4}]$ term. The critical exponents in this limit become $\eta_\psi = \mathcal{O}(1/N_f)$, $\eta_{\varphi}= \epsilon$, and $1/\nu = 2- \epsilon +\mathcal{O}(1/N_f)$, which are the usual large-$N$ values.

Our analysis focused on the physical dimension $d=2$, and we therefore analytically continued  only the one-loop integrals to non-integer dimensions. Had we extended the gamma-matrix algebra as well, near $d=3$ the two real and one imaginary Pauli matrices would, respectively, be replaced by three real and one imaginary $4\times4$ gamma-matrices \cite{herbuttimereversal}, with the rest of calculation remaining the same. The sole effect of this replacement would be that our parameter  $N_f \rightarrow 2 N_f$. 
Similar modification happens if we consider two dimensional two-component $N_{f}M$ copies of the Weyl fermions instead of the Majorana fermions.
For instance, the critical values of this $N_{f}$ would behave as $N_{f,c2} \approx N$ in the large $N$ limit.

We found that a large number of fermion flavors $N_{f, c2} \approx 2N-2$ is needed to find a stable fixed point and the continuous transition in the tensor GNY theory. The situation resembles the one in the scalar Higgs electrodynamics \cite{herbutbook} where similarly a large number of complex order parameters is required to overcome fluctuations of the gauge field and the concomitant runaway RG flow. It is now understood, however, that in this \cite{halperin, kolnberger, tesanovic1, tesanovic2, higgs} and similar problems \cite{klebanov1, klebanov2, janssen, roscher} this conclusion is often an artifact of the first-order epsilon expansion, and that higher-order corrections may significantly modify the value of $N_{f,c2}$. If we write
\begin{equation}
N_{f, c2} = \sum_{n=0}^\infty f_{n} (N) \epsilon ^n,
\end{equation}
then only the first term $f_{0}(N)$ has been determined here. It would be interesting to find the sign and the size of the next-order correction $f_1 (N)$, which follows from two-loops \cite{steudtner, jack}
as it may reduce the critical value $N_{f,c2}$ for the appearance of the continuous transition. One is particularly interested in the GNY theory at the GN-relevant value of $N_f=1$, which, however, may be difficult to access by the epsilon expansion. Conformal bootstrap \cite{erramilli, reehorst} seems to be more promising for this purpose.

Finally, we suspect that the emergence of the critical fixed point in GNY theories with matrix order parameters at large number of fermions may be a rather generic feature of this class of field theories. Indeed, a similar  large-$N_f$ fixed point has been found in the theory for the matrix order parameter which is in the adjoint representation of the $\text{SO}(4)$ \cite{herbutso4, uetrecht, kaul}. The mechanism at work is as described right below Eq.~\eqref{eq:lambda2_fixed}, and follows from the fact that the double-trace interaction coupling becomes negligible when $N_f \gg 1$.

\section{Acknowledgement}

This work was supported by the NSERC of Canada.

\end{document}